# Effects of Chemical Pressure on Superconductivity in Electrochemically Intercalated (TMA)$_y$Fe$_2$(Se$_{1-x}$S$_x$)$_2$ (TMA = Tetramethylammonium)

Nadine Lammer, Dominik Werhahn, Leonard Moritz and Dirk Johrendt*

Department Chemie

Ludwig-Maximilians-Universität München

Butenandtstrasse 5–13 (D), 81377 München (Germany)

Email Address: johrendt@lmu.de

The *β*-modification of FeSe, which has an *anti*-PbO type structure, achieves superconductivity at 8K without external doping or pressure and exhibits a nematic phase, which has been crucial for studies of unconventional pairing mechanisms. Although the critical temperature ($T_c$) in FeSe increases significantly with applied pressure, intercalation, and in thin films, the effect of replacing selenium with sulfur in FeSe$_{1-x}$S$_x$ has remained unclear. To investigate the effects of chemical pressure, we have synthesized FeSe$_{1–x}$S$_x$ crystals (up to $x$ = 0.52) and intercalated them with tetramethylammonium ions (TMA$^+$). Our results show that both $T_c$ and unit cell volume decrease linearly with sulfur content in both host and intercalated (TMA)$_y$Fe$_2$(Se$_{1-x}$S$_x$)$_2$. The unexpected common rate of normalized decrease of $T_c$ suggests that chemical pressure affects the electron-doped intercalates in the same way as the host compounds. This result highlights the unique role of chemical pressure and electron doping in tuning superconductivity in FeSe systems and provides new insights into their complex behavior.

## 1 Introduction

The *β*-modification of iron selenide has an *anti*-PbO type structure, consisting of stacked layers of edgesharing FeSe$_{4/4}$ tetrahedra [1], and is a superconductor with a critical temperature ($T_c$) of 8 K [2]. FeSe stands out among iron-based superconductors because superconductivity is achieved without doping or applying pressure, unlike in iron arsenide compounds [3]. It displays no magnetic ordering but symmetrybreaking antiferromagnetic correlations with spatial anisotropy, known as the nematic phase [4]. The relatively low $T_c$ of FeSe increases under pressure to 37K [5], reaches up to 46K by intercalation [6, 7], and can rise to as high as 99K in thin films [8]. The remarkable tuneability of the physical properties has been the subject of numerous studies [9, 10], with the coexistence of the nematic and superconducting phases and their intertwining being of particular interest for the understanding of unconventional superconductivity [11, 12, 13]. It has been shown that substituting selenium with sulfur in FeSe$_{1-x}$S$_x$ gradually suppresses nematicity [14, 15]. However, the reported effects on the change in $T_c$ are contradictory. One might expect that the reduction in volume, due to the incorporation of smaller sulfur atoms, would produce effects similar to those caused by external pressure. This is the case in BaFe$_2$(As$_{1-x}$P$_x$)$_2$, where superconductivity is induced by either external or chemical pressure [16]. According to literature until 2021, sulfur substitution in FeSe$_{1-x}$S$_x$ did not increase $T_c$ [17], until Sun *et al.* reported the solid solution FeSe$_{1-x}$S$_x$ ($x$ = 0–1) with an increase of $T_c$ up to 37K at $x \approx 0.35$, and a phase diagram similar to that of FeSe under external pressure [18]. The claimed synthetic twist was to cure the iron defects in Fe$_{1-\delta}$Se$_{1-x}$S$_x$ by additionally treating the crystals with excess iron powder in an alkaline medium.

As mentioned above, $T_c$ is not only increased by compression but also by expansion of the FeSe layers by intercalation of ions or molecules. Values as high as 46 K have been achieved when charged molecules and/or metal cations are incorporated, resulting in electron doping of the FeSe layers. Several studies with intercalated FeSe under pressure have shown that the critical temperature initially decreases, and a second maximum occurs at higher pressure [19, 20, 21]. In contrast, nothing is known about the effect of chemical pressure on FeSe intercalates yet. This is a particularly interesting question because the chemical pressure acts explicitly on the layers themselves and not on the layer spacing.

To address this question, we have synthesized crystals of the host compounds FeSe$_{1-x}$S$_x$ ($x$ = 0–0.52;1) by chemical vapor transport and then electrochemically intercalated them with tetramethylammonium ions (TMA$^+$). We find that the volumes and critical temperatures of the host FeSe$_{1-x}$S$_x$ and the intercalated TMA$_y$Fe$_2$(Se$_{1-x}$S$_x$)$_2$ both decrease at a common linear rate up to $x \approx 0.5$, suggesting that the chemical pressure acts unexpectedly similarly on the electron-doped intercalated compounds as on the host material.

## 2   Results and Discussion

FeSe$_{1-x}$S$_x$ host compounds ($x$ = 0–0.52;1) were synthesized with the chemical vapor transport (CVT) method [22, 23] and characterized by powder X-ray diffraction, elemental analyses, and magnetic susceptibility measurements (see Supporting Information). All samples crystallize in the tetragonal *anti*-PbO type structure (space group *P4/nmm*) with atom positions close to those of FeSe and FeS, respectively. They exhibit superconductivity with critical temperatures between approximately 9 and 4 K. Tetragonal FeS does not form with the CVT method and was prepared from iron and sulfur under hydrothermal conditions [24].

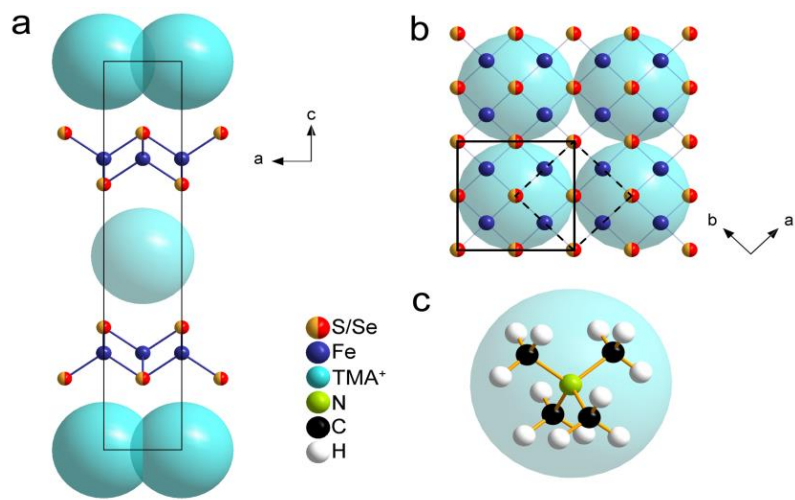

Figure 1: Crystal structure of $(TMA)_yFe_2(Se_{1-x}S_x)_2$ in the space group $I4/mmm$, according to Rendenbach et al. [25]. Large turquoise spheres represent TMA+ cations with a diameter of ≈ 5.5 Å [26, 27]. (a) View along the $b$ axis. (b) The $\sqrt{2}a \times \sqrt{2}a$ supercell (solid black line) with perfectly fitting TMA+ cations, which occupy half the sites. The unit cell is marked with a dashed line. (c) Ball-and-stick representation of a TMA+ molecule.

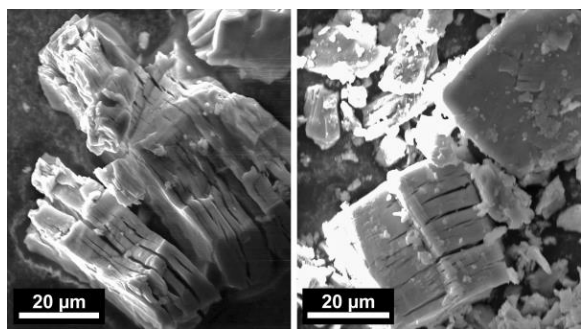

Figure 2: SEM images of two $(TMA)_yFe_2(Se_{1-x}S_x)_2$ samples. Left: $(TMA)_{0.38}Fe_2Se_{1.60}S_{0.40}$ ($x = 0.20$); Right: $(TMA)_{0.38}Fe_2Se_{0.96}S_{1.04}$ ($x = 0.52$).

The electrochemical intercalation of TMA+ ions into the van der Waals gap of the host material expands the interlayer distance and shifts the layers by [½, ½, 0]. This layer configuration corresponds to the $ThCr_2Si_2$ structure (space group $I4/mmm$), which is characteristic of superconducting iron arsenides such as $BaFe_2As_2$ [28]. Figure 1 shows the crystal structure of $TMA_{0.5}Fe_2Se_2$, which was adopted from Rendenbach et al. [25]. TMA+ ions are too large to fit within a unit cell with a 3.8 Å lattice parameter (Figure 1a), so a $\sqrt{2}a \times \sqrt{2}a$ supercell is employed, where TMA+ occupies half of the available sites (Figure 1b). This superstructure is undetectable by X-ray diffraction due to the weak scattering of light atoms and the rotational disorder of the TMA+ molecules. According to energy dispersive X-ray spectroscopy (EDX) measurements (Table S2), the sulfur contents in the $FeSe_{1-x}S_x$ layers remain unchanged upon intercalation. Rietveld refinements are not reliable for accurately determining the composition because the platelet-like shape of the crystals induces preferred orientation, which distorts the diffraction intensities. The layered morphology of the crystals is also evident in the scanning electron microscopy (SEM) images (Figure 2). The amount of intercalated $TMA^+$ cations was determined by CHNS elemental analysis with an accuracy of ±0.05 (Table S1). With increasing sulfur incorporation, the amount of the $TMA^+$ cations may decrease slightly. However, the change is within two times the error.

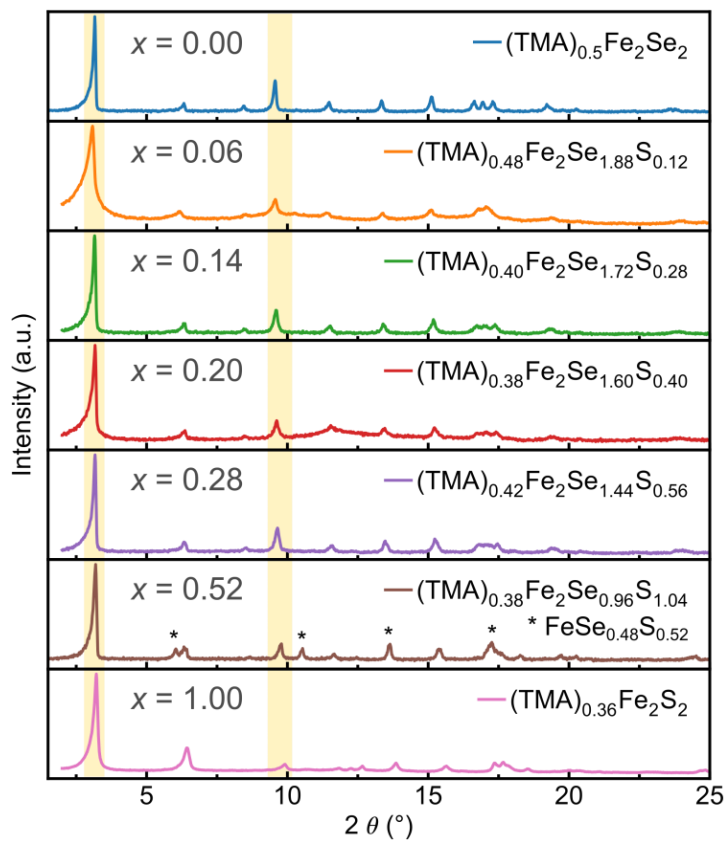

Figure 3: Powder X-ray diffraction patterns of the various $(TMA)_y Fe_2(Se_{1-x}S_x)_2$ compounds. Marked in yellow are the (002) reflection (left), representing the elongation of the $c$ axis due to the successful intercalation, and the (006) reflection together with the (103) reflection (right), visualizing the slight shift to larger diffraction angles with decreasing cell size. For $x = 0.52$, around 28wt% precursor is still present in the sample.

Figure 3 shows the powder X-ray diffraction patterns. A slight shift in the reflections (highlighted in yellow) indicates the incorporation of sulfur into the host lattice. The 'curved' background observed at around 10°$2\theta$ for $x=0.06$ and at around 12°$2\theta$ for $x=0.20$ is attributed to amorphous impurities, probably due to trace amounts of mercury. Rietveld refinements (Figure S1) suggest a 100% intercalation yield within the detection limits of the method (~1wt%). The main results are summarized in Table 1; further details can be found in the Supporting Information. The presence of about 28 wt% of the precursor in the sample with $x = 0.52$ is probably due to insufficient contact between mercury and the precursor.

Table 1: $T_c$ and lattice parameters of (TMA)$_y$Fe$_2$(Se$_{1-x}$S$_x$)$_2$ with agreement indices of the Rietveld refinements.

| Sample | x | $T_c$ | a (Å) | c (Å) | V (Å$^3$) | $R_p$ | $R_{wp}$ | $R_{exp}$ | $R_{bragg}$ |
| --- | --- | --- | --- | --- | --- | --- | --- | --- | --- |
| (TMA)$_{0.5}$Fe$_2$Se$_2$[25] | 0 | 43 K | 3.8585(2) | 20.377(3) | 303.4(1) | 1.940 | 2.789 | 0.815 | 1.269 |
| (TMA)$_{0.48}$Fe$_2$Se$_{1.88}$S$_{0.12}$ | 0.06 | 41 K | 3.8253(10) | 20.663(19) | 302.4(3) | 2.167 | 2.797 | 1.406 | 0.449 |
| (TMA)$_{0.40}$Fe$_2$Se$_{1.72}$S$_{0.28}$ | 0.14 | 39 K | 3.8367(5) | 20.201(7) | 297.36(13) | 4.997 | 6.293 | 6.416 | 1.599 |
| (TMA)$_{0.38}$Fe$_2$Se$_{1.60}$S$_{0.40}$ | 0.20 | 37 K | 3.8270(8) | 20.121(11) | 294.7(2) | 3.193 | 4.096 | 3.162 | 1.059 |
| (TMA)$_{0.42}$Fe$_2$Se$_{1.44}$S$_{0.56}$ | 0.28 | 35 K | 3.8154(4) | 20.128(7) | 293.01(12) | 4.669 | 5.907 | 5.738 | 1.869 |
| (TMA)$_{0.38}$Fe$_2$Se$_{0.96}$S$_{1.04}$ | 0.52 | 28 K | 3.7673(6) | 20.140(8) | 285.84(15) | 7.449 | 9.525 | 9.591 | 1.820 |
| (TMA)$_{0.36}$Fe$_2$S$_2$ | 1 | - | 3.7038(5) | 19.823(7) | 271.94(12) | 5.746 | 7.737 | 2.481 | 4.197 |

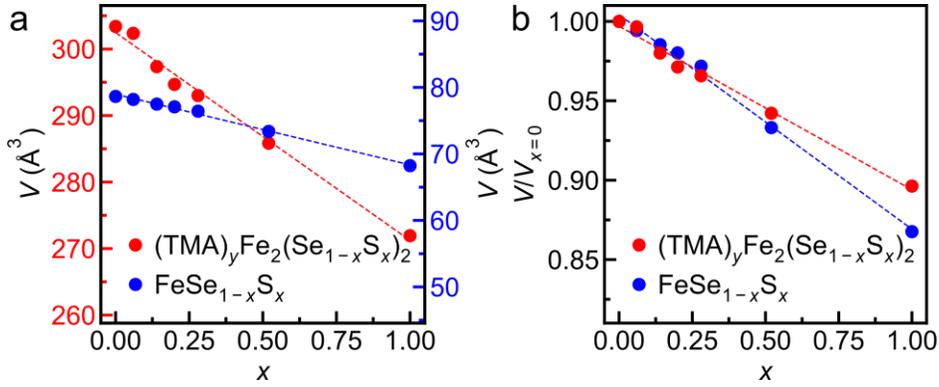

Figure 4: Change of unit cell volume with the sulfur content for (TMA)$_y$Fe$_2$(Se$_{1-x}$S$_x$)$_2$ (red) and the precursors FeSe$_{1-x}$S$_x$ (blue). (a) Absolute volumes. (b) Normalized volumes $V/V_{x=0}$.

As sulfur incorporation increases, the lattice parameters decrease as expected, reflecting the smaller ionic radius of sulfur ($r$ = 1.84 Å) compared to selenium ($r$ = 1.98 Å) [29]. The volumes of both the host FeSe$_{1-x}$S$_x$ and the intercalated (TMA)$_y$Fe$_2$(Se$_{1-x}$S$_x$)$_2$ decrease linearly with rising sulfur content (Figure 4a). Although the absolute volume changes appear more pronounced in the intercalated samples, the normalized volume changes in the host and intercalate are quite similar (Figure 4b). This suggests that the substitution of selenium by sulfur in the host lattice exerts the same effect on the FeSe$_{1-x}$S$_x$ layers as in the intercalated compounds.

The structural (nematic) transition of FeSe is rapidly suppressed by sulfur incorporation in FeSe$_{1-x}$S$_x$ [14, 15] and is absent in TMA$^+$-intercalated compounds (Figure S3). While (TMA)$_{0.5}$Fe$_2$Se$_2$ deintercalates to FeSe at 200°C [25], our high-temperature X-ray diffraction measurements (Figure S4) show that (TMA)$_{0.36}$Fe$_2$S$_2$ decomposes at 150°C, with FeS changing from the tetragonal to the hexagonal modification [30, 31, 32]. Thus, the intercalation of TMA$^+$ into tetragonal FeS is irreversible.

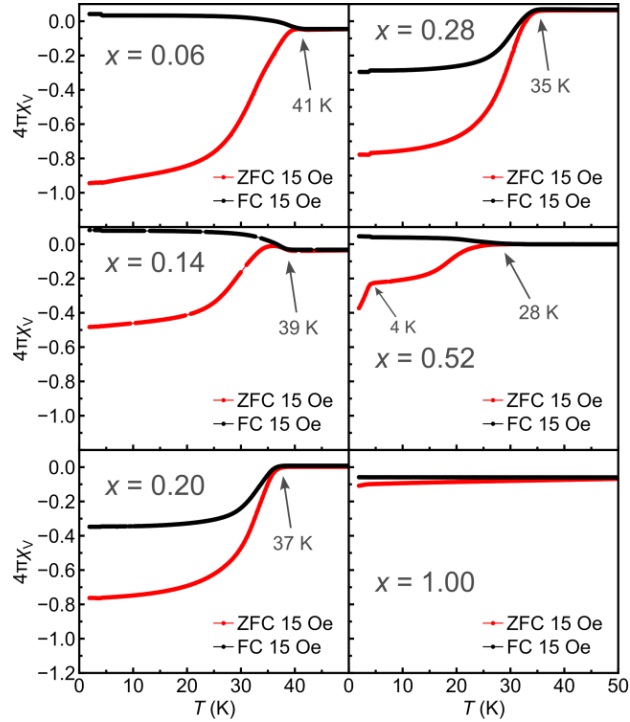

Figure 5: ZFC (red) and FC (black) curves of $(TMA)_y Fe_2(Se_{1-x}S_x)_2$ samples at 15 Oe. Upturns in some of the FC branches (see Figure S2) may be manifestations of a paramagnetic Meissner effect [33] or due to interference of ferromagnetic impurities with the diamagnetic Meissner signal.

Low-field susceptibility measurements taken after zero-field cooling are shown in Figure 5. Samples with $x$ = 0.06–0.28 show superconductivity with critical temperatures decreasing from 41 K to 35 K. Superconducting volume fractions range from 0.5 to 0.95 for compositions up to $x$ = 0.28, indicating bulk superconductivity. At $x$ = 0.52 the superconducting volume fraction decreases to about 20% and the critical temperature $T_c$ drops to 28 K. The additional drop at 4K is attributed to residual precursor and/or mercury. No superconducting signal was detected for $(TMA)_{0.36}Fe_2S_2$, suggesting that intercalation suppresses the low-temperature superconductivity of FeS (∼4 K). This contrasts with the precursor material, where $T_c$ initially increases to 9–15 K at $x$ = 0.1–0.2 before decreasing with further substitution to finally reach $T_c \approx$ 4 K for $FeSe_{0.5}S_{0.5}$ and FeS [34, 35, 24].

Isothermal magnetization curves of $(TMA)_y Fe_2(Se_{1-x}S_x)_2$ samples at 2K show the typical 'butterfly' pattern of hard type-II superconductors with large upper critical fields $H_{c2}$ (Figure 6). We do not see flux jumps as observed in $(TMA)_{0.5}Fe_2Se_2$ [25]. The magnetization pattern changes particularly strongly from $x$ = 0.28 to $x$ = 0.52, where the critical fields become significantly smaller. Finally, non-superconducting $(TMA)_{0.36}Fe_2S_2$ shows a very weak signal as an S-shaped curve, which is caused by traces of ferromagnetic impurities stemming from the hydrothermal synthesis of the FeS precursor.

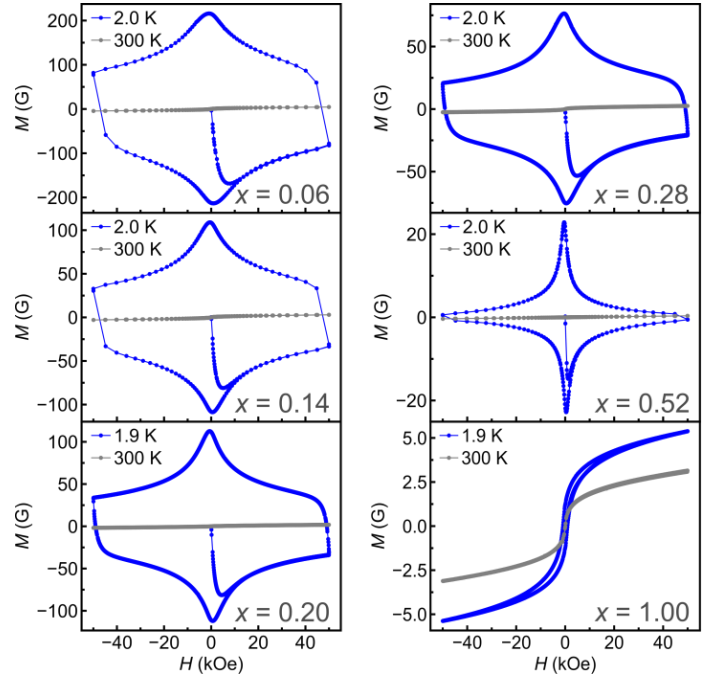

Figure 6: Magnetization isotherms of the $(TMA)_y Fe_2(Se_{1-x}S_x)_2$ samples at ∼2 K (blue) and 300 K (gray).

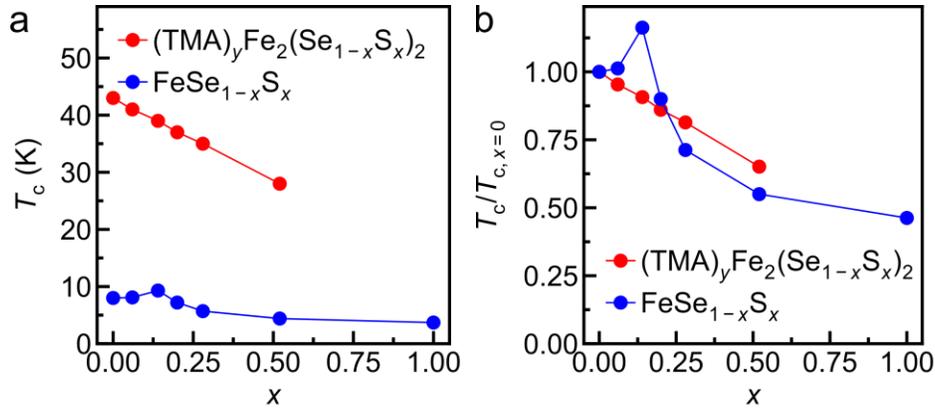

Figure 7: Superconducting critical temperatures ($T_c$) of the intercalated $(TMA)_y Fe_2(Se_{1-x}S_x)_2$ (red) and $FeSe_{1-x}S_x$ host (blue) compounds. (a) Absolute $T_c$ values. (b) Normalized values $T_c/T_{c,x=0}$.

The critical temperature of the $TMA^+$-intercalated samples decreases linearly with sulfur content, from 43 K at $x = 0$ to 28 K at $x = 0.52$, where superconductivity weakens in terms of critical field and volume fraction (Figures 7a and 6). As shown in Figure 7b, the normalized critical temperatures unexpectedly show similar rates of decrease in both the host and intercalated compounds. This suggests that the chemical pressure exerted on the layers proportionally scales down the higher $T_c$ generated by electron doping.

However, at larger sulfur contents where the host compounds remain superconducting [17], the chemical pressure seems to completely suppress superconductivity. Unlike some related FeSe intercalates under physical pressure [21], we do not observe a second increase in $T_c$. Since we did not reach concentrations above $x$ = 0.52, however, we cannot entirely rule out the possibility of another $T_c$ maximum.

## 3  Conclusion

We have demonstrated that substituting sulfur for selenium in (TMA)$_y$Fe$_2$(Se$_{1-x}$S$_x$)$_2$ applies a form of chemical pressure, scaling the superconducting critical temperature at a rate closely matching that observed in the host compounds FeSe$_{1-x}$S$_x$. It is reasonable to assume that the significantly higher $T_c$ of the intercalated samples is primarily due to electron doping provided by the TMA$^+$ ions, as intercalation with charge-neutral molecules has been shown to yield materials with either low $T_c$ [36, 37] or non-superconducting properties [38]. The unexpected proportionality observed in our study suggests that TMA$^+$ intercalates of host compounds with $T_c$ enhancements under chemical pressure, as reported by Sun *et al.* [18], may achieve very high critical temperatures - potentially even exceeding 77K. However, this would apply only to electronically undoped host compounds. However, it may be possible that the high-$T_c$ FeSe$_{1-x}$S$_x$ compounds reported by Sun are actually doped with either iron or sodium between the layers due to the excess presence of these elements during synthesis. Our findings suggest that chemical pressure, electron doping, and even minimal compositional changes can significantly impact superconductivity in FeSe and its intercalates. Given the preparative and analytical challenges of controlling these variations, further research is essential to deepen understanding and enable consistent formation of superconductivity in iron selenides.

## 4  Experimental Section

*Synthesis*:

The synthesis procedure reported by Rendenbach *et al.* [25] was modified for sulfur substitution. The $β$-FeSe$_{1-x}$S$_x$ precursors were synthesized using chemical vapor transport [23]. According to the nominal compositions in Table S3, Fe powder (Chempur, 99.9%), Se powder (Chempur, 99.999%) and S powder (Sigma-Aldrich, 99.99%) (1g in total) were ground together with 7.75g AlCl$_3$ (Thermo Scientific, 99.985%) and 2.25g dried KCl (Grüssing, 99.5%). The mixture was sealed under vacuum in a glass ampoule (10cm in height, 6cm in diameter) and placed into a 2-Gradient Split Tube Furnace (HZS2G Model, 425mm, 3216 controllers) by Carbolite Gero. The samples were heated up to 390°C (bottom) and 240°C (top) with a dwell time of 8 days (or 20 days for the sample with nominal composition $Fe_{1.1}Se_{0.3}S_{0.7}$). After opening the ampoule to air, the sample was washed with water and ethanol and dried *in vacuo*.

The FeS precursor was hydrothermally synthesized according to Lai *et al.* [24]. In a Teflon-liner (50mL), 1.40g Fe powder (Chempur, 99.9%) and 6.00g $Na_2S \cdot 9 H_2O$ (Sigma-Aldrich, ≥99.99%) were mixed with 20mL degassed distilled water. It was subsequently sealed in a stainless-steel autoclave under argon atmosphere and heated at 140°C for 6 days. After cooling, the precipitate was collected by centrifugation and washed with distilled water (3x 10mL), and ethanol (1x 5mL). Small amounts of unreacted Fe were removed using a magnet during washing. Afterwards, the product was dried *in vacuo*.

Subsequently, $FeSe_{1-x}S_x$ host crystals were electrochemically intercalated with tetramethylammonium cations ($TMA^+$). A tungsten rod served as the anode, while an amalgamated copper spoon attached to a platinum wire served as the cathode. To ensure good electrical contact, a drop of mercury filled the spoon [39] on which the host crystals were dispersed. The cathode was immersed in a saturated solution of tetramethylammonium iodide (TMAI, Sigma-Aldrich, 99%) dissolved in 100mL dry dimethylformamide (DMF, Thermo Scientific, 99.8% extra dry) inside of an electrolysis cell under argon atmosphere. A voltage of three volts was applied between the electrodes for three days. The $(TMA)_yFe_2(Se_{1-x}S_x)_2$ products were washed with dry DMF and subsequently dried *in vacuo*.

*Analytics*:

To confirm the formation of the tetragonal structure of the $FeSe_{1-x}S_x$ precursors, powder X-ray diffraction (pXRD) was performed on a Huber G670 diffractometer (Cu-$K_{\alpha 1}$, $\lambda$ = 1.54056 Å; Guinier Camera). The composition was determined by inductively coupled plasma optical emission spectroscopy (ICP-OES), elemental CHNS analysis, and EDX. The intercalated products were investigated using a Stoe STADI P diffractometer (Ag-$K_{\alpha 1}$, $\lambda$ = 0.55941 Å). CSD 2394652–2394657 contain the supplementary crystallographic data. These data can be obtained free of charge from FIZ Karlsruhe via www.ccdc.cam.ac.uk/structures. The composition was confirmed by elemental CHNS analysis and EDX. The superconducting properties were investigated by measuring the magnetic susceptibility using a vibrating sample magnetometer (VSM) unit incorporated into a physical property measurement system (PPMS) (Quantum Design Inc.). Moreover, low-temperature powder X-ray diffraction (LT-pXRD) was performed on a Huber G670 diffractometer (Co-K$_{\alpha 1}$, $\lambda$ = 1.78892 Å; Guinier Camera) equipped with a low-temperature device, while high-temperature powder X-ray diffraction (HT-pXRD) was carried out on Stoe STADI P diffractometer (Ag-K$_{\alpha 1}$, $\lambda$ = 0.55941 Å) fitted with a graphite furnace. SEM and EDX measurements were carried out on a Carl Zeiss EVO-MA 10 electron microscope with an integrated EDX system (Bruker X-Flash 410-M detector).

More detailed information about the analytical methods is described in the Supporting Information.

**Acknowledgements**

This work was funded by the German Research Foundation (DFG), Grant Nr. JO257/7-2. We thank Dr. Dieter Rau for measuring the HT-pXRD patterns.

# Effects of Chemical Pressure on Superconductivity in Electrochemically Intercalated (TMA)$_y$Fe$_2$(Se$_{1-x}$S$_x$)$_2$ (TMA = Tetramethylammonium)

Nadine Lammer, Dominik Werhahn, Leonard Moritz, and Dirk Johrendt

## Supporting Information

## 1 Powder X-ray Diffraction

The polycrystalline samples were analyzed using powder X-ray diffraction (pXRD) on a Stoe STADI P diffractometer (Ag-K$_{\alpha1}$, $\lambda$ = 0.55941 Å; Ge(111) monochromator; Si as external standard) with rotating capillaries (0.3mm in diameter, Hilgenberg GmbH) and a Dectris MYTHEN 1K strip detector in Debye-Scherrer geometry in transmission. The collected data were processed using the WinXPOW software,[1] while Rietveld refinements were carried out with the Topas software package.[2] The crystal structure was visualized by the program Diamond.[3]

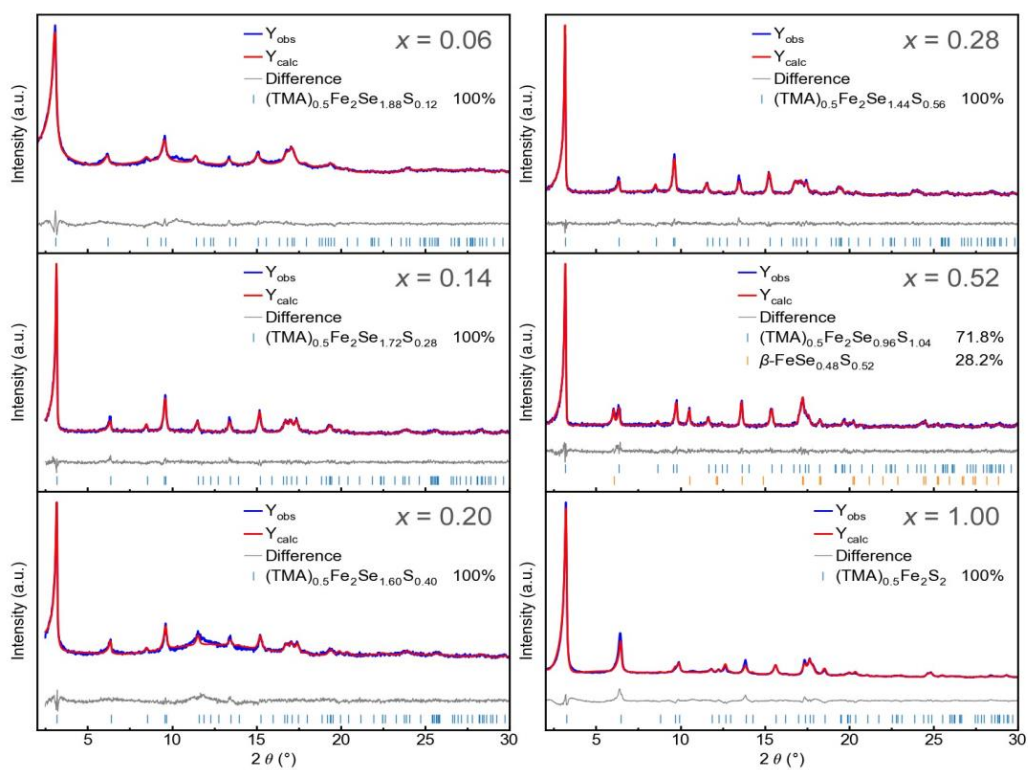

**Figure S1:** Rietveld refinements of the (TMA)$_y$Fe$_2$(Se$_{1-x}$S$_x$)$_2$ samples with the measured data (blue), Rietveld fit (red), and difference curve (gray).

Figure S1 shows the Rietveld refinements of the compounds in the space group $I4/mmm$. For $x$ = 0.52, the remaining precursor is also included in the refinement. For the Rietveld refinements, the amount of nitrogen was set to 0.5, like in $(TMA)_{0.5}Fe_2Se_2$, since the organic part only contributes to a small degree to the intensities in the diffraction patterns and the carbon atoms are not accounted for. Moreover, the occupancy of selenium and sulfur was not refined due to the preferred orientation of the platelet-like structure, which influences the reflection intensities.

## 2 Composition

The amount of intercalated $TMA^+$ was determined by CHNS elemental analysis by calculating the N:S ratio and assuming the Fe:Se:S ratio of the precursor (Table S1). The lack of standard deviation is due to the single determination of the samples. For $x$ = 0.52, the amount of $TMA^+$ inside the compound was calculated based on the amount of intercalation product according to the Rietveld refinement. The N:C:H ratios of around 1:4:12 are in accordance with the $C_4H_{12}N^+$ cation, only slightly deviating, showing the integrity of the $TMA^+$ cation. The amount of intercalated $TMA^+$ cations is in the range of 0.36 to 0.48 cations per $Fe_2Se_2$ unit, corresponding to a site occupancy factor (SOF) of 0.36 - 0.48 at the $2a$ site.

**Table S1:** Compositions by CHNS elemental analysis. N:C:H ratio normalized to N = 1, and S:N ratio normalized to S = 1.

| Sample | $x$ | CHNS | | | | | Composition |
|---|---|---|---|---|---|---|---|
| | | N | C | H | S | N | |
| $(TMA)_yFe_2Se_{1.88}S_{0.12}$ | 0.06 | 1 | 4.12 | 14.7 | 1 | 4.34 | $(TMA)_{0.48}Fe_2Se_{1.88}S_{0.12}$ |
| $(TMA)_yFe_2Se_{1.72}S_{0.28}$ | 0.14 | 1 | 3.92 | 12.2 | 1 | 1.42 | $(TMA)_{0.40}Fe_2Se_{1.72}S_{0.28}$ |
| $(TMA)_yFe_2Se_{1.60}S_{0.40}$ | 0.20 | 1 | 4.46 | 19.0 | 1 | 0.96 | $(TMA)_{0.38}Fe_2Se_{1.60}S_{0.40}$ |
| $(TMA)_yFe_2Se_{1.44}S_{0.56}$ | 0.28 | 1 | 4.20 | 14.9 | 1 | 0.76 | $(TMA)_{0.42}Fe_2Se_{1.44}S_{0.56}$ |
| $(TMA)_yFe_2Se_{0.96}S_{1.04}$ | 0.52 | 1 | 4.26 | 15.3 | 1 | 0.36 | $(TMA)_{0.38}Fe_2Se_{0.96}S_{1.04}$ |
| $(TMA)_yFe_2S_2$ | 1 | 1 | 5.29 | 16.6 | 1 | 0.18 | $(TMA)_{0.36}Fe_2S_2$ |

The compositions of the products were confirmed by energy-dispersive X-ray spectroscopy (EDX) and only show slight deviations (Table S2). The largest difference is for $(TMA)_{0.40}Fe_2Se_{1.72}S_{0.28}$ ($x$ = 0.14), whose precursor has the highest discrepancy as well. This might be due to a wider range of incorporated sulfur in different crystallites. The EDX measurements were conducted on a Carl Zeiss EVO-MA 10 electron microscope with an integrated EDX setup (Bruker X-Flash 410-M detector). The software SmartSEM[4] provided an interface to control the detectors (SE and BSE), while the EDX data acquisition and processing were carried out using the software QUANTAX 200[5].

Table S2: Fe:Se:S ratios from EDX normalized to Se + S = 2.

| Composition | EDX | | |
|---|---|---|---|
| | Fe | Se | S |
| $(TMA)_{0.48}Fe_2Se_{1.88}S_{0.12}$ | 2.40(12) | 1.88(2) | 0.12(2) |
| $(TMA)_{0.40}Fe_2Se_{1.72}S_{0.28}$ | 2.36(14) | 1.62(6) | 0.38(6) |
| $(TMA)_{0.38}Fe_2Se_{1.60}S_{0.40}$ | 2.42(12) | 1.58(4) | 0.42(4) |
| $(TMA)_{0.42}Fe_2Se_{1.44}S_{0.56}$ | 2.24(12) | 1.42(10) | 0.58(10) |
| $(TMA)_{0.38}Fe_2Se_{0.96}S_{1.04}$ | 2.16(8) | 0.90(8) | 1.10(8) |
| $(TMA)_{0.36}Fe_2S_2$ | 2.32(18) | 0 | 2 |

## 3 Magnetic Measurements

Magnetic measurements were carried out on a vibrating sample magnetometer (VSM) unit incorporated into a physical property measurement system (PPMS) (Quantum Design Inc.). Zero-field cooling (ZFC) and field cooling (FC) curves were measured between 2K and 60K at a magnetic field of 15Oe. Magnetization isotherms were recorded at 2K and 300K between ±50kOe. The software PPMS MultiVu[6] provided an interface for the data acquisition.

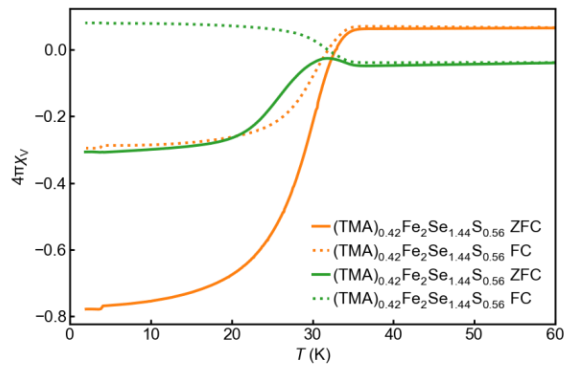

**Figure S2:** Zero-field cooling (ZFC) and field cooling (FC) curves of $(TMA)_{0.42}Fe_2Se_{1.44}S_{0.56}$ ($x$ = 0.28) at 15Oe, indicating $T_c$ above the upturn.

The effect of the positive FC curve and the upturn in the ZFC curve is not always reproducible, as seen in Figure S2. The critical temperature $T_c$ can be determined at the temperature above the upturn in the ZFC curve.

## 4 Temperature Dependent Diffraction Data

Low-temperature powder X-ray diffraction (LT-pXRD) measurements were carried out on a Huber G670 diffractometer (Co-K$_{\alpha 1}$, $\lambda$ = 1.78892 Å; Ge(111) monochromator, Guinier Camera) with flat sample holders, equipped with a low-temperature device cooled by a closed cycle He cryostat. The temperature of the measurements ranged between 10K and 300K.

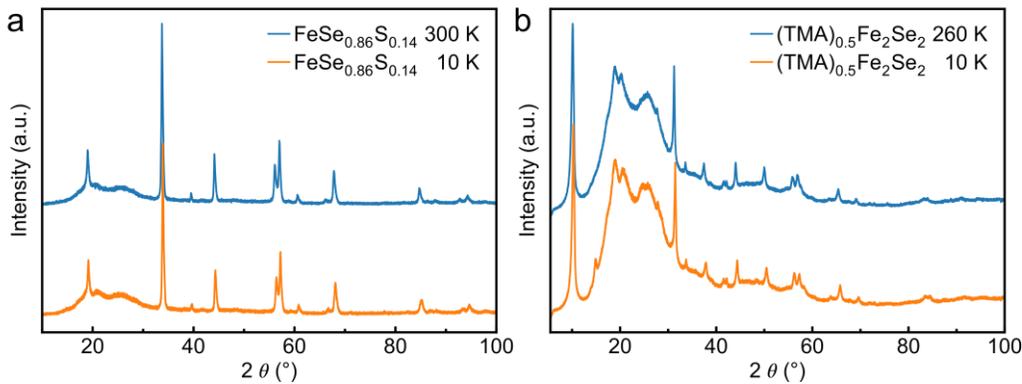

**Figure S3:** Low-temperature diffraction patterns of (a) FeSe$_{0.86}$S$_{0.14}$, and (b) (TMA)$_{0.5}$Fe$_2$Se$_2$, revealing no phase transition in either the precursor or the intercalated product. High background due to the usage of grease to fixate the powder sample.

The LT-pXRD patterns of the precursor FeSe$_{0.86}$S$_{0.14}$ and the intercalated (TMA)$_{0.5}$Fe$_2$Se$_2$ are shown in Figure S3. No nematic phase transition is visible in either sample. The large amorphous background for (TMA)$_{0.5}$Fe$_2$Se$_2$ stems from using grease to fixate the sample for the measurement.

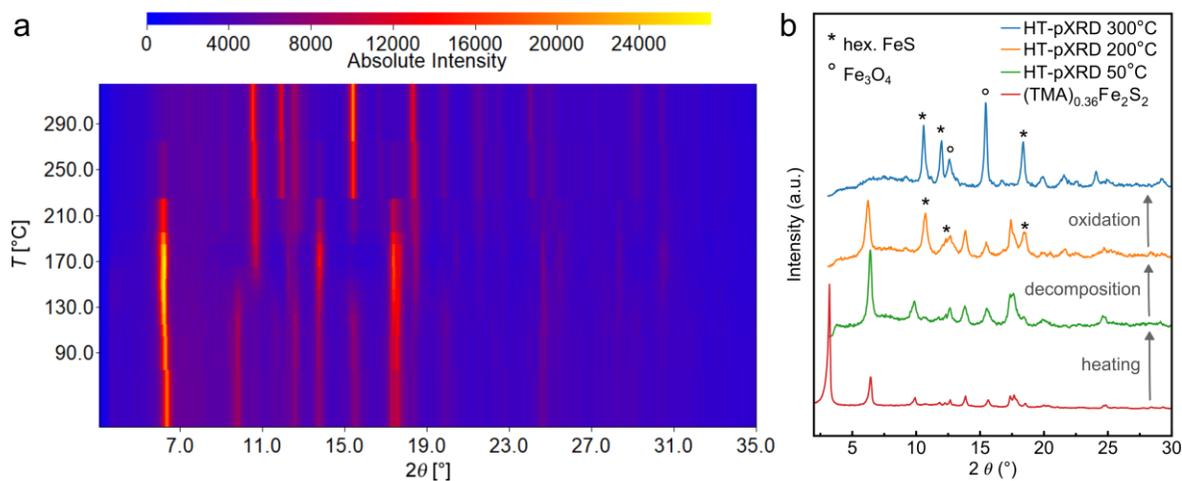

**Figure S4:** High-temperature powder diffraction patterns of (TMA)$_{0.36}$Fe$_2$S$_2$. (a) Diffraction patterns between 50 and 300K as top view. (b) Diffraction patterns at 50°C (green), 200°C (orange), and 300°C (blue), with an RT diffraction pattern for comparison (red), showing the decomposition and successive oxidation during the heating. The latter reveals exposure to air or humidity during the measurement.

The high-temperature powder X-ray diffraction (HT-pXRD) measurement was performed on a Stoe STADI P diffractometer (Ag-K$_{\alpha 1}$, $\lambda$ = 0.55941 Å; Ge(111) monochromator, IP-PSD detector in Debye-Scherrer geometry) with rotating quartz capillaries (0.5mm in diameter, Hilgenberg GmbH) and a graphite furnace.

Figure S4 displays the HT-pXRD of (TMA)$_{0.36}$Fe$_2$S$_2$. At 150°C the sample starts to decompose to hexagonal FeS, showing that the intercalated iron sulfide cannot be deintercalated to tetragonal FeS. The successive oxidation is likely due to exposure to air or humidity during the measurement, caused by an improperly sealed capillary or water residues inside the capillary.

## 5  Characterization of the Precursors

The precursors were investigated by powder X-ray diffraction. These were performed on a Huber G670 diffractometer (Cu-K$_{\alpha 1}$, $\lambda$ = 1.54056 Å; Ge(111) monochromator, Guinier Camera) with flat sample holders. Rietveld refinements were performed using the Topas software package.[2]

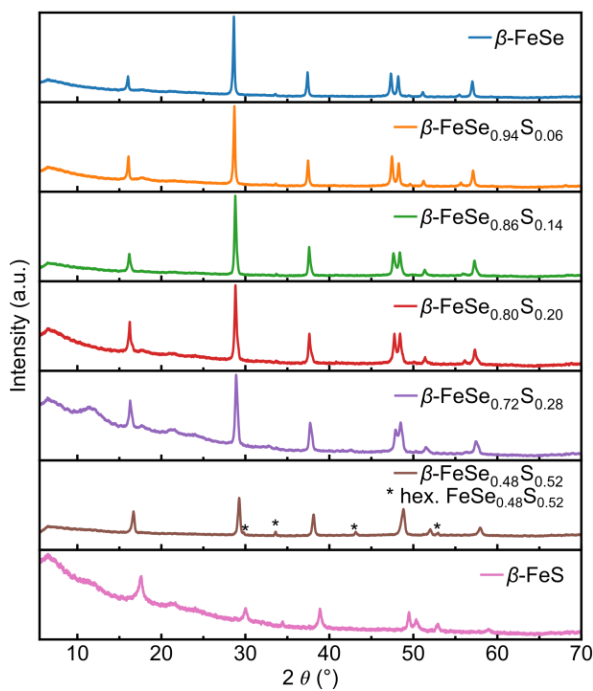

**Figure S5:** Powder diffraction patterns of the various synthesized FeSe$_{1-x}$S$_x$ precursors. For $x$ = 0.52, around 6wt% of a hexagonal side phase is present.

All precursors crystallize in the tetragonal space group *P*4/*nmm* (Figure S5). No impurities could be determined by pXRD, except for sample FeSe$_{0.48}$S$_{0.52}$, where a hexagonal side phase (*P*6$_3$/*mmc*) is present (5.9wt%). This is likely due to the reaction temperature being too high for the sulfur-rich tetragonal phase.[7–9]

The composition of the FeSe$_{1-x}$S$_x$ precursors was investigated by inductively coupled plasma optical emission spectroscopy (ICP) for Fe and Se in combination with CHNS elemental analysis for S, and EDX measurements for all three elements (Table S3). The ICP/CHNS elemental analyses were singledetermined. Hence, no standard deviation is listed. The sulfur amount and, therefore, the composition was determined by averaging over the two different methods since ICP/CHNS has the drawback of combining two different analytical methods, while EDX only averages over a smaller sample size. The two different methods only diverge slightly, the largest being Δ$x$ = 0.06 for FeSe$_{0.86}$S$_{0.14}$. Incorporating sulfur into the structure seems unpreferable since the amount of sulfur in the samples is less than the nominal amount. This is in accordance with Mizuguchi *et al.*[10] for solid-state syntheses, and Pachmayr[11] for hydrothermal syntheses, where a large thiourea excess has to be used.

**Table S3:** Nominal compositions, Fe:Se:S ratios determined by ICP and CHNS, as well as EDX, and determined compositions of FeSe$_{1-x}$S$_x$ compounds. All ratios normalized to Se + S = 1.

| Nominal | ICP&CHNS | EDX | Determined |
|---------|----------|-----|------------|

| Composition | Fe | Se | S | Fe | Se | S | Composition |
|---|---|---|---|---|---|---|---|
| Fe$_{1.1}$Se$_{0.9}$S$_{0.1}$ | 1.02 | 0.94 | 0.06 | 1.11(6) | 0.95(1) | 0.05(1) | FeSe$_{0.94}$S$_{0.06}$ |
| Fe$_{1.1}$Se$_{0.8}$S$_{0.2}$ | 1.04 | 0.83 | 0.17 | 1.12(5) | 0.89(3) | 0.11(3) | FeSe$_{0.86}$S$_{0.14}$ |
| Fe$_{1.1}$Se$_{0.7}$S$_{0.3}$ | 1.09 | 0.79 | 0.21 | 1.15(6) | 0.82(2) | 0.18(2) | FeSe$_{0.80}$S$_{0.20}$ |
| Fe$_{1.1}$Se$_{0.5}$S$_{0.5}$ | 1.07 | 0.71 | 0.29 | 1.09(3) | 0.73(4) | 0.27(4) | FeSe$_{0.72}$S$_{0.28}$ |
| Fe$_{1.1}$Se$_{0.3}$S$_{0.7}$ | 1.03 | 0.48 | 0.52 | 1.08(5) | 0.49(5) | 0.51(5) | FeSe$_{0.48}$S$_{0.52}$ |
| FeS | 1.14 | 0 | 1 | 1.04(7) | 0 | 1 | FeS |

**Table S4:** $T_c$ and lattice parameters of the FeSe$_{1-x}$S$_x$ with agreement indices of the Rietveld refinements.

| Sample | $T_c$ | $a$ (Å) | $c$ (Å) | $V$ (Å$^3$) | $R_p$ | $R_{wp}$ | $R_{exp}$ | $R_{bragg}$ |
|---|---|---|---|---|---|---|---|---|
| FeSe | 8 K | 3.7731(1) | 5.5239(2) | 78.641(4) | 1.431 | 2.042 | 1.026 | 2.476 |
| FeSe$_{0.94}$S$_{0.06}$ | 8.1 K | 3.7686(1) | 5.5051(2) | 78.187(4) | 1.250 | 1.705 | 1.040 | 1.501 |
| FeSe$_{0.86}$S$_{0.14}$ | 9.3 K | 3.7610(1) | 5.4782(4) | 77.487(7) | 1.430 | 2.202 | 0.938 | 1.485 |
| FeSe$_{0.80}$S$_{0.20}$ | 7.2 K | 3.7578(1) | 5.4586(4) | 77.080(7) | 1.138 | 1.631 | 1.072 | 0.370 |
| FeSe$_{0.72}$S$_{0.28}$ | 5.7 K | 3.7505(2) | 5.4330(6) | 76.421(12) | 1.186 | 1.656 | 0.643 | 0.512 |
| FeSe$_{0.48}$S$_{0.52}$ | 4.4 K | 3.7221(1) | 5.2967(3) | 73.381(6) | 1.031 | 1.459 | 0.927 | 1.108 |
| FeS | 3.7 K | 3.6805(2) | 5.0367(8) | 68.229(14) | 0.992 | 1.255 | 1.221 | 0.572 |

Table S4 lists the lattice parameters of the precursors obtained from the Rietveld refinements and the corresponding agreement indices. Since sulfur has a smaller ionic radius than selenium, the lattice parameters decrease in all dimensions. The occupancies of selenium and sulfur were not refined because of the platelet-like form of the crystallites, inducing a preferred orientation.

As seen in Figure S6, all precursors are superconducting. With increasing sulfur incorporation, $T_c$ decreases after a slight increase until it reaches a value of 4K at $x \approx 0.5$. It was impossible to increase the sulfur amount further between $0.52 < x < 1$. These findings are in accordance with the literature.[10,12] Tetragonal iron sulfide is also superconducting at 4 K, as found by Lai *et al.*[13].

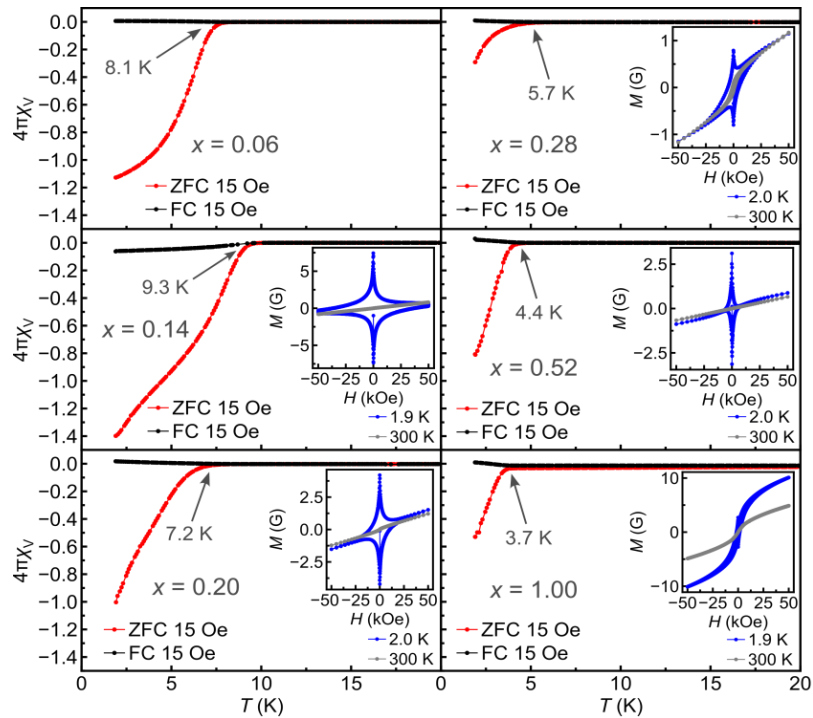

**Figure S6:** Zero-field cooling (red) and field cooling (black) curves of the synthesized FeSe$_{1-x}$S$_x$ precursors at 15 Oe. (inlets) Magnetization isotherms at ~2 K (blue) and 300 K (gray).